\documentstyle[prl,aps]{revtex}
\draft
\begin{document}
\title{Ground State Properties of Anderson Impurity in a Gapless Host}
\author{Valery I. Rupasov\cite{ea}}
\address{Department of Physics, University of Toronto, Toronto,
Ontario, Canada M5S 1A7\\
and\\
Landau Institute for Theoretical Physics, Moscow, Russia}
\date{\today}
\maketitle
\begin{abstract}
Using the Bethe ansatz method, we study the ground state
properties of a $U\to\infty$ Anderson impurity in a
``gapless'' host, where a density of band states vanishes
at the Fermi level $\epsilon_F$ as $|\epsilon-\epsilon_F|$.
As in metals, the impurity spin is proven to be screened
at arbitrary parameters of the system. However, the impurity
occupancy as a function of the bare impurity energy is shown
to acquire novel qualitative features which demonstrate a
nonuniversal behavior of the system. The latter explains why
the Kondo screening is absent (or exists only at quite a large
electron-impurity coupling) in earlier studies based on
scaling arguments.
\end{abstract}


The physics of ``gapless'' dilute magnetic alloys, where
an effective density of band electron states varies near
the Fermi level as $|\epsilon-\epsilon_F|^r$, $r>0$, has
been attracting a significant theoretical interest. Using
poor-man's scaling for the spin-$\frac{1}{2}$ Kondo model,
Withoff and Fradkin \cite{WF} have predicted that the
Kondo effect in gapless systems takes place only if an
effective electron-impurity coupling exceeds some critical
value; otherwise, the impurity decouples from the electron
band. Numerical renormalization group (RG) calculations,
large-$N$ studies, and quantum Monte Carlo simulations
\cite{BH,CF,CJ,I,GBI,BPH} have confirmed this prediction
and revealed a number of additional intriguing features
of the physics of magnetic impurities in unconventional
Fermi systems.

In a conventional metallic system with (i) a linear
dispersion of electrons near the Fermi level,
$\epsilon(k)=v_F(k-k_F)$, and (ii) an energy independent
hybridization, basic ``impurity'' models are exactly solved
by the Bethe ansatz (BA) \cite{TW,AFL,S,H}. It has recently
been shown \cite{R} also that integrability of the
$U\to\infty$ nondegenerate and degenerate Anderson
models is not destroyed by a nonlinear dispersion  of
particles and an energy dependent hybridization, but
it becomes only hidden \cite{RS}. The approach developed
has allowed us to study \cite{R2} the thermodynamic and
ground state properties of an Anderson impurity embedded
in a BCS superconductor, and can be used to obtain an exact
solution of the Kondo problem in other unconventional
systems.

In this Letter, we report an exact BA solution of a model
describing a $U\to\infty$ Anderson impurity embedded in a
gapless host. The model is diagonalized by BA at the
arbitrary density of band states $\rho(\epsilon)$ and
hybridization $t(\epsilon)$. In the RG approach, the
physics of the system is assumed to be governed only by
an effective electron-impurity coupling $\Gamma(\epsilon)=
\rho(\epsilon)t^2(\epsilon)$ rather than by separate forms
of $\rho(\epsilon)$ and $t(\epsilon)$. However, to derive
thermodynamic BA equations, one has to specify separate
forms of an effective electron-impurity coupling and an
inverse dispersion of band states $k(\epsilon)$. While an
effective coupling determines the electron-impurity and
effective electron-electron scattering amplitudes, an
inverse dispersion accounts for the spatial behavior of
electron wave functions, and naturally enters BA equations
via periodic boundary conditions imposed on eigenfunctions
of the system. The physics of the system is thus governed
by both an effective electron-impurity coupling $\Gamma(\epsilon)$
and an inverse dispersion of band electrons $k(\epsilon)$
\cite{R3}.

Here, we treat the case of an energy independent hybridization,
$t(\epsilon)= t=\mbox{constant}$, so that an energy dependence
of an effective coupling $\Gamma(\epsilon)=2\Gamma\rho(\epsilon)$,
where $2\Gamma=t^2$, is determined only by a nonlinear band
dispersion. We assume also a simple form for the density of
states of a gapless host,
\begin{equation}
\rho(\epsilon)=\frac{|\epsilon|^r}{|\epsilon|^r+\beta^r},
\end{equation}
where the energy $\epsilon$ is taken relative to the Fermi
value $\epsilon_F$.  The parameter $\beta$ characterizes
the size of region with an unconventional behavior of
$\rho(\epsilon)$. At $\beta=0$, the model reduces to the
metallic Anderson model. If $\beta$ exceeds essentially a
band half width $D$, $\beta\gg D$, one obtains the density
of states $\rho(\epsilon)\sim|\epsilon|^r$.  To derive
thermodynamic BA equations one has to fix the power $r$
in Eq. (1). The magnitude of $r$ is a key factor in
determining the spectrum of the system in terms of Bethe
excitations. In this Letter, we focus on the simplest case
$r=1$, which is however of particular physical interest
\cite{WF,BH,CF,CJ,I,GBI,BPH}.

In the BA technique, the spectrum of elementary excitations
of the metallic $U\to\infty$ Anderson model is described in
terms of unpaired charge excitations, charge complexes, and
spin excitations including bound spin complexes \cite{TW,AFL,S,H}.
Since the ground state of the system is composed only of
charge complexes (a charge complex contains one spin wave
and two charge excitations) carrying no spin, the Kondo
effect takes place: the impurity spin vanishes at zero
temperature.

In the gapless model described above, the structure of
the spectrum is shown to preserve basic characteristic
features of the metallic version. In particular, the
ground state of the system is still composed only of
charge complexes carrying no spin. Therefore, as in
metals, the Kondo screening of the impurity spin takes
place at arbitrary parameters of the system, that
contradicts dramatically to results of earlier studies
\cite{WF,BH,CF,CJ,I,GBI,BPH}.

However, the behavior of the impurity occupancy $n_d$ as
a function of the bare impurity level energy $\epsilon_d$
is drastically changed compared to a metal host. At positive
values of $\epsilon_d$, the impurity occupancy is still given
by the standard formulae \cite{TW}, where the renormalized
impurity level energy $\epsilon^*_d=\epsilon_d+
\frac{\Gamma}{\pi}\ln{\frac{D}{\Gamma}}+\beta$ contains now
the parameter $\beta$. The mixed-valence regime shrinks: the
impurity occupancy quickly grows from $n_d\approx 0$ at
$\epsilon_d=0$ to $n_d=1$ (precisely!) at
$\epsilon_d\leq -\Gamma^2/4\beta$. In the Kondo (local-moment)
regime, $n_d=1$ and it does not depend on $\epsilon_d$. Only in
the empty-level regime ($\epsilon_d>0$) the impurity occupancy
is a universal function of the renormalized impurity energy
$\epsilon^*_d$ rather than a function of the bare parameters
of the model.

The behavior of the impurity occupancy in the mixed-valence
and Kondo regimes is not universal that manifests nonuniversal
properties of a gapless system in contrast to a metallic one.
This explains why the Kondo screening is absent (or exists
only at quite a large electron-impurity coupling) in earlier
studies based on scaling arguments.

We start with the Hamiltonian of the nondegenerate Anderson
model rewritten in terms of the Fermi operators
$c^\dagger_\sigma(\epsilon)$ ($c_\sigma(\epsilon)$) which
create (annihilate) an electron with spin
$\sigma=\uparrow,\downarrow$ in an $s$-wave state of energy
$\epsilon$,
\begin{mathletters}
\begin{equation}
H=H_c+H_d+H_h.\\
\end{equation}
Here
\begin{eqnarray}
H_c&=&\sum_{\sigma}\int_{-D}^{D}\frac{d\epsilon}{2\pi}\epsilon
c^\dagger_\sigma(\epsilon)c_\sigma(\epsilon)\\
H_d&=&\epsilon_d\sum_{\sigma} d^\dagger_\sigma d_\sigma +
Ud^\dagger_\uparrow d_\uparrow d^\dagger_\downarrow d_\downarrow\\
H_h&=&\sum_{\sigma}\int_{-D}^{D}d\epsilon\sqrt{\Gamma(\epsilon)}
[c^\dagger_\sigma(\epsilon)d_\sigma+d^\dagger_\sigma c_\sigma(\epsilon)].
\end{eqnarray}
are the conduction band, impurity and hybridization terms,
respectively. All notation in Eqs. (2) are standard. An
electron localized in an impurity orbital with the energy
$\epsilon_d$ is described by the Fermi operators $d_\sigma$.
The electron energies and momenta are taken relative to the
Fermi values, which are set to be equal to zero. The
integration over the energy variable $\epsilon$ is
restricted by the band half width $D$. In what follows,
we assume that $D$ is the largest parameter on the
energy scale, $D\to\infty$. In the energy representation,
the effective particle-impurity coupling
$\Gamma(\epsilon)=\rho(\epsilon)t^2(\epsilon)$ combines
the density of band states, $\rho(\epsilon)=dk/d\epsilon(k)$,
and the energy dependent hybridization $t(\epsilon)$.

In the limit of a large Coulomb repulsion in an impurity
orbital, $U\gg D$, eigenvalues of the model (2) with the
arbitrary inverse dispersion $k(\epsilon)$ and effective
coupling $\Gamma(\epsilon)$ are found from the following
BA equations \cite{R}:
\end{mathletters}
\begin{mathletters}
\begin{eqnarray}
e^{ik_jL}\frac{h_j-\epsilon_d/2\Gamma-i/2}
{h_j-\epsilon_d/2\Gamma+i/2}&=&\prod_{\alpha=1}^{M}
\frac{h_j-\lambda_\alpha-i/2}{h_j-\lambda_\alpha+i/2}\\
\prod_{j=1}^{N}\frac{\lambda_\alpha-h_j-i/2}
{\lambda_\alpha-h_j+i/2}
&=&-\prod_{\beta=1}^{M}
\frac{\lambda_\alpha-\lambda_\beta-i}{\lambda_\alpha-\lambda_\beta+i}
\end{eqnarray}
where $N$ is the total number of electrons on an interval of
size $L$ and $M$ is the number of electrons with spin ``down''.
The eigenenergy $E$ and the $z$ component of total spin
of the system are found to be
\begin{equation}
E=2\Gamma\sum_{j=1}^{N}\omega_j,\;\;\;S^z=\frac{N}{2}-M.
\end{equation}

In Eqs. (3), $\omega=\epsilon/2\Gamma$ is a dimensionless energy,
and $k_j=k(\omega_j)$ are charge excitation momenta. From Eq. (1),
one easily obtains
\end{mathletters}
\begin{mathletters}
\begin{equation}
\frac{k}{2\Gamma}=\left\{
\begin{array}{ll}
\omega+\delta\ln{(1-\frac{\omega}{\delta})},&\omega<0\\
\omega-\delta\ln{(1+\frac{\omega}{\delta})},&\omega>0,
\end{array}
\right.
\end{equation}
where $\delta=\beta/2\Gamma$. The BA equations (3) are
quite similar to those in the conventional Anderson model
\cite{TW,AFL,S,H}. The only but very essential difference
is a nonlinear energy dependence of momenta and charge
``rapidities'' $h_j=h(\omega_j)$. At arbitrary $\rho(\epsilon)$
and $t(\epsilon)$, the rapidity
$h(\epsilon)=(\epsilon-\epsilon_d)/\Gamma(\epsilon)$. In
our model of a gapless host, where $t^2(\epsilon)=2\Gamma$
and $\rho(\epsilon)$ is defined as in Eq. (1) with $r=1$,
\begin{equation}
h(\omega)=\left(\omega-\frac{\epsilon_d}{2\Gamma}\right)
\frac{|\omega|+\delta}{|\omega|}+\frac{\epsilon_d}{2\Gamma}.
\end{equation}

As in the metallic Anderson model, in the thermodynamic limit
spin rapidities $\lambda_\alpha$, $\alpha=1,\ldots,M$ are
grouped into bound spin complexes of size $n$,
\end{mathletters}
\begin{equation}
\lambda_\alpha^{(n,j)}=\lambda_\alpha+\frac{i}{2}(n+1-2j),\;\;\;
j=1,\ldots,n.
\end{equation}
Apart from unpaired charge excitations with real rapidities
$h_j$, the system spectrum contains also charge complexes,
in which two charge excitations with complex rapidities bound
to a spin wave with a real rapidity $\lambda_\alpha$,
\begin{equation}
h_\alpha^{(\pm)}(\omega)=\lambda_\alpha\pm\frac{i}{2}.
\end{equation}

In the absence of an impurity term [the second term
in l. h. s. of Eq. (3a)], BA equations of ``impurity''
models describe a free host in terms of interacting
Bethe particles with an arbitrary rapidity $h(\omega)$.
Introducing an impurity fixes an expression for $h(\omega)$,
and fixes thus the spectrum of Bethe excitations of a
host. For instance, in the theory of metallic magnetic
alloys, the finite-$U$ and $U\to\infty$ Anderson impurities
require different descriptions of the same host with
the different spectra of Bethe excitations. In a gapless
system, the impurity energy $\epsilon_d$ is involved not
only in the impurity term but also in the expression for
the rapidity $h(\omega)$. The magnitude of $\epsilon_d$
thus determines a particle-impurity scattering phase and
dictates also an appropriate choice of the Bethe spectrum
of a host.

It is instructive to start our analysis of the ground state
properties of the system with the simplest case $\epsilon_d=0$.
The complex energies of charge excitations of a charge complex
are then found to be
\begin{equation}
\omega_\pm(\lambda)=\left\{
\begin{array}{ll}
\lambda+\delta\pm\frac{i}{2},& \lambda<-\delta\\
\lambda-\delta\pm\frac{i}{2},& \lambda>\delta
\end{array}
\right.
\end{equation}
In contrast to the metallic case, the spectrum of charge
complexes contains thus the gap of size $2\delta$. The
existence of the gap should lead to essential changes
in the thermodynamics of the system. However, it can be
shown from the thermodynamic Bethe ansatz equations that
the renormalized energies of unpaired charge excitations
and spin complexes are still positive in the absence of an
external magnetic field. Therefore, as in the conventional
Anderson model, the ground state of the system is composed
of charge complexes with negative renormalized energies
only. The upper edge of filled states, $Q_G$, is now given by
$Q_G=Q_A-\delta$, where $Q_A=-\frac{1}{2\pi}\ln{\frac{D}{\Gamma}}$
corresponds to the metallic Anderson model. The occupancy
of the impurity level is also given by the standard BA formulae
\cite{TW}, where in the expression for the renormalized impurity
level energy $\epsilon^*_d$ one needs only to replace $Q_A$ by
$Q_G$ to get $\epsilon^*_d=\frac{\Gamma}{\pi}\ln{\frac{D}{\Gamma}}
+\beta$. Thus, the impurity occupancy at $\epsilon_d=0$ is
essentially decreased compared to the metallic case.

At $\epsilon_d\neq 0$, the complex energies of charge
excitations of a complex are found from the equation
\begin{equation}
\omega_\pm^2-(\lambda+\delta\pm\frac{i}{2})\omega_\pm
+\delta\frac{\epsilon_d}{2\Gamma}=0.
\end{equation}
Since we study here only the ground state of the system, we
may restrict our consideration to the solutions with the negative
real part of energies, $\mbox{Re}\,\omega<0$. Then, a solution
$\omega_\pm(\lambda)=x(\lambda)\pm i y(\lambda)$ is given by
\begin{mathletters}
\begin{equation}
x(\lambda)=\frac{1}{2}\left[\mu-u(\mu)\right];\;\;\;
y(\lambda)=\frac{1}{2}\left[\frac{1}{2}-v(\mu)\right],
\end{equation}
where
\begin{eqnarray}
u&=&\frac{1}{\sqrt{2}}\left[\mu^2-b+\sqrt{(\mu^2-b)^2+\mu^2}\right]^{1/2}\\
v&=&\frac{\mbox{sgn}\mu}{\sqrt{2}}
\left[-\mu^2+b+\sqrt{(\mu^2-b)^2+\mu^2}\right]^{1/2}
\end{eqnarray}
and $\mu=\lambda+\delta$. The behavior of this solution is
governed by the parameter
$b=\frac{1}{4}+4\frac{\epsilon_d}{2\Gamma}\delta$ \cite{N}.

Let us consider first the case of positive $\epsilon_d$,
and hence $b>\frac{1}{4}$.  Then, as in the case $\epsilon_d=0$,
the function $x(\lambda)$ is negative only at $\lambda<-\delta$.
The ground state of the system is still composed of charge
complexes filling all the states from $\lambda=-D/2\Gamma$
to $\lambda=Q_G$. The impurity occupancy $n_d$ is governed
by the well known formulae with insignificant corrections
related to small deviations of the function $x(\lambda)$ from
the linear behavior.

Significant changes in the behavior of the system occur at
negative $\epsilon_d$. At negative $b$ ($\epsilon_d/2\Gamma<
-1/16\delta$), the function $x(\lambda)$ is negative at all
$\lambda\in(-\infty,\infty)$. In contrast both to the metallic
model and to the gapless system with positive $\epsilon_d$,
the bare energy of charge complexes,
$\xi_0(\lambda)=4\Gamma x(\lambda)$, is now monotonically
increasing negative function at all $\lambda$. Therefore,
in the ground state of the system charge complexes fill out
all allowed states on the $\lambda$ axis. The density of
states of charge complexes $\sigma(\lambda)$ is found from
the continuous limit \cite{TW,AFL,S,H} of Eqs. (3),
\end{mathletters}
\begin{equation}
\frac{1}{2\pi}\frac{dq(\lambda)}{d\lambda}+
\frac{1}{L}a(\lambda-\frac{\epsilon_d}{2\Gamma})=
\sigma(\lambda) =
\int_{-\infty}^{\infty}d\lambda' a(\lambda-\lambda')
\sigma(\lambda'),
\end{equation}
where $q(\lambda)=k_-(\lambda)+k_+(\lambda)$ is the
momentum of the charge complexes, and $a(x)=[\pi(x^2+1)]^{-1}$.
As usual, the function $\sigma(\lambda)$ is divided into the
host and impurity parts, $\sigma(\lambda)=\sigma_h(\lambda)
+L^{-1}\sigma_i(\lambda)$. The occupancy of the impurity level,
$n_d$, is then given by
\begin{equation}
n_d=2\int_{-\infty}^{\infty}d\lambda\sigma_i(\lambda),
\end{equation}
where the impurity density of states is found from the
equation
\begin{equation}
a(\lambda-\frac{\epsilon_d}{2\Gamma})=\sigma_i(\lambda)
+\int_{-\infty}^{\infty}d\lambda' a(\lambda-\lambda')
\sigma_i(\lambda').
\end{equation}
Since unpaired charge excitations and spin complexes
are absent in the ground state of the system, the
impurity spin vanishes. Thus, as in the metallic
Anderson system, the Kondo effect takes place at
an arbitrary particle-impurity coupling.

Solving Eq. (11), we immediately find $n_d=1$. Thus, at
$\epsilon_d<-\Gamma^2/4\beta$ the impurity level is
entirely filled out and its occupancy does not depended
on a position of the impurity energy with respect to the
Fermi level. Thus, the behavior of the impurity occupancy
in the gapless host is not described by a universal function
of the renormalized impurity energy $\epsilon^*_d$ but it
depends essentially on the bare parameters of the model
$\epsilon_d$ and $\beta$. Making use of the terminology
of the Anderson model, we will call this regime with the
entirely filled impurity level the Kondo (or local-moment)
regime, despite it disappears in the limit $\beta\to 0$,
where our model reduces to the conventional Anderson model.
However, for quite large $\delta$, and even at $\delta\leq 1$,
the Kondo regime describes the system's behavior almost at
all negative $\epsilon_d$, except a very narrow region near
the Fermi level, where $0<b<\frac{1}{4}$.

The region $0<b<\frac{1}{4}$ corresponds in our case to
the mixed-valence regime, where the impurity occupancy
is changed from $n_d\approx 0$ in the empty-level regime
at $\epsilon_d\geq 0$ to $n_d=1$ precisely in the Kondo
regime at $\epsilon_\leq-\Gamma^2/4\beta$. The bare
energy of charge complexes $\xi_0(\lambda)=4\Gamma x(\lambda)$
in this regime is negative at all $\lambda$, except the point
$\lambda=-\delta$, where $x(\lambda)=0$. The renormalized
energy of charge complexes at zero temperature is found
from the thermodynamic BA equation
\begin{equation}
\xi(\lambda)=4\Gamma x(\lambda)
-\int_{-\infty}^{Q_1}d\lambda' a(\lambda-\lambda')\xi(\lambda')
-\int_{Q_2}^{\infty}d\lambda' a(\lambda-\lambda')\xi(\lambda'),
\end{equation}
where $Q_1$ and $Q_2$ are defined as the zeroes of $\xi(\lambda)$,
$\xi(Q_1)=\xi(Q_2)=0$. This equation, as well as a corresponding
equation for the density of states $\sigma(\lambda)$, is hardly
solved analytically, and a numerical analysis is required.

In the metallic limit, $\beta\to 0$, the lower edge of the
mixed-valence regime in the gapless system is shifted to
$-\infty$. Correspondingly, the mixed-valence regime of the
gapless system is extended to the conventional mixed-valence
($0<n_d<1$) and Kondo ($n_d\approx 1$) regimes of the Anderson
impurity in a metallic host.

In summary, using the BA we have studied the ground state
properties of an $U\to\infty$ Anderson impurity embedded
in a gapless host with an energy independent hybridization
and the density of band states given in Eq. (1) with the
power $r=1$. As in the metallic version, the ground state
of the system has been shown to be composed of charge
complexes only. Since each complex contains two charge
excitation and one spin wave, the total spin of a complex
equals zero. Therefore, at zero temperature the impurity
spin vanishes, and the Kondo effect takes place at arbitrary
parameters of the model.

However, the appearance of extra energy parameter $\beta$,
characterizing the size of region with an unconventional
behavior of the density of band electron states leads to
significant reconstruction of the density of states of charge
complexes in the ground state of the system. This reconstruction
results in significant changes in the behavior of the impurity
occupancy as a function of the bare impurity level energy.
The empty-level ($n_d\approx 0$) and Kondo ($n_d=1$) regimes
are extended to almost all positive and negative values of
$\epsilon_d$, respectively. While the mixed-valence regime
($0<n_d<1$) is squeezed to a narrow region
$-\Gamma^2/4\beta<\epsilon_d<0$.

Only in the empty-level regime the impurity occupancy is
a universal function of the renormalized impurity energy
$\epsilon^*_d=\epsilon_d+\frac{\Gamma}{\pi}\ln{\frac{D}{\Gamma}}
+\beta$, which contains the parameter $\beta$. The behavior
of $n_d$ in the mixed-valence and Kondo regimes is not
universal that demonstrates nonuniversal properties of
the system.

The latter explains why the Kondo screening is absent -
or exists only at quite a large electron-impurity
coupling - in earlier studies \cite{WF,BH,CF,CJ,I,GBI,BPH}
based on scaling arguments. Nevertheless, by using both
poor-man's scaling and numerical RG calculations,
Gonzalez-Buxton and Ingersent have derived the behavior
of the impurity occupancy which is qualitatively close
to the picture described above. However, in their studies
\cite{I,GBI} the particle-hole symmetry is a key factor
in determining the low-temperature physics of the system,
while in the BA solution, the ground state properties are
obviously insensitive to this symmetry.

In the BA analysis, the power $r$ must be fixed. The
spectrum of the system is determined by the functions
$k(\omega)$ and $h(\omega)$, and it is essentially
different at different $r$. At $r\neq 1$, the spectrum
of Bethe excitations is enriched that could result in
qualitatively novel physical properties of the system.
Although it seems very difficult, if not impossible,
to propose any scenario of destroying the Kondo screening
of an integrable Anderson impurity.

I thank S. John for stimulating discussions.


\begin{references}

\bibitem[*]{ea}
Electronic address: rupasov@physics.utoronto.ca

\bibitem{WF}
D. Withoff and E. Fradkin, \prl {\bf 64}, 1835 (1990).

\bibitem{BH}
L. S. Borkowski and P. J. Hirschfeld, \prb {\bf 46},
9274 (1992).

\bibitem{CF}
C. R. Cassanello and E. Fradkin, \prb {\bf 53}, 15079 (1996);

\bibitem{CJ}
K. Chen and C. Jayaprakash, J. Phys.: Condens. Matter,
{\bf 7}, L491 (1995).

\bibitem{I}
K. Ingersent, \prb {\bf 54}, 11936 (1996).

\bibitem{GBI}
C. Gonzalez-Buxton and K. Ingersent, \prb {\bf 54}, 15614 (1996);
preprint cond-mat/9803256.

\bibitem{BPH}
R. Bulla, Th. Pruschke, and A. C. Hewson, J. Phys.: Condens. Matter
{\bf 9}, 10463 (1997).

\bibitem{TW}
A. M. Tsvelick and P. B. Wiegmann, Adv. Phys. {\bf 32}, 453 (1983).

\bibitem{AFL}
N. Andrei, K. Furuya, and J. H. Lowenstein, \rmp {\bf 55}, 331 (1983).

\bibitem{S}
P. Schlottmann, Phys. Rep. {\bf 181}, 1 (1989).

\bibitem{H}
A. C. Hewson, {\em The Kondo Effect to Heavy Fermions}
(Cambridge University Press, Cambridge, 1993).

\bibitem{R}
V. I. Rupasov, Phys. Lett. {\bf 237A}, 80 (1997).

\bibitem{RS}
V. I. Rupasov and M. Singh, J. Phys. A {\bf 29}, L205 (1996);
\prl {\bf 77}, 338 (1996); \pra {\bf 54}, 3614 (1996).

\bibitem{R2}
V. I. Rupasov, \prl {\bf 80}, (1998).

\bibitem{R3}
V. I. Rupasov, preprint cond-mat/9807231; \prb {\bf 58},
No. 17 (Rapid Communications) (1998).

\bibitem{N}
At $b>\frac{1}{4}$, the second root of Eq. (8)
$x^{(+)}(\lambda)=\frac{1}{2}[\mu+u]$ also satisfies
the condition $x^{(+)}(\lambda)<0$ at $\lambda<-\delta$,
that allows us to construct more complex charge complexes.
However, as unpaired charge excitations and spin complexes,
they do not contribute to the ground state of the system.
The same scenario occurs in the finite-$U$ metallic Anderson
model \cite{TW}.

\end{references}
\end{document}